\documentclass[useAMS,usenatbib]{mn2e}

\usepackage{epsfig}
\usepackage{amsmath}
\usepackage{amssymb}
\usepackage{natbib}
\usepackage{threeparttable} 

\usepackage{epstopdf}

\newcommand{\Msun}{\mathrm{M}_{\odot}}
\newcommand{\lum}{\mathrm{erg~s}^{-1}}

\newcommand{\flux}{\mathrm{erg~cm}^{-2}~\mathrm{s}^{-1}}
\newcommand{\fluence}{\mathrm{erg~cm}^{-2}}
\newcommand{\cnts}{\mathrm{counts~s}^{-1}}

\newcommand{\nh}{\mathrm{cm}^{-2}}

\newcommand{\xmmu}{XMMU~J1747}
\newcommand{\xmmuname}{XMMU~J174716.1--281048}
\newcommand{\igr}{IGR~J17464--2811}
\newcommand{\rxh}{1RXH~J173523.7--354013}

\newcommand{\chan}{\textit{Chandra}}

\newcommand{\swift}{\textit{Swift}}
\newcommand{\xmm}{\textit{XMM-Newton}}

\newcommand{\inte}{\textit{INTEGRAL}}

\def \mnras {MNRAS}
\def \apj {ApJ}
\def \apjs {ApJS}
\def \apjl {ApJL}
\def \aap {A\&A}

\def \nar {NewAR}

\title[X-ray burst from \xmmuname]{\swift\ detection of an intermediately long X-ray burst from the very faint X-ray binary \xmmuname}
\author[N. Degenaar, R. Wijnands \& R. Kaur]
{N. Degenaar$^{1}$\thanks{e-mail: degenaar@uva.nl}, R. Wijnands$^{1}$ and R. Kaur$^{1,2}$\\
$^{1}$Astronomical Institute ``Anton Pannekoek", University of Amsterdam, Postbus 94249, 1090 GE, Amsterdam, the Netherlands\\
$^{2}$Department of Physics, University of Wisconsin, Milwaukee, WI 53201-0431, USA
}

\begin{document}

\date{Accepted 2011 April 8.  Received 2011 April 8; in original form 2011 March 21.}

\pagerange{\pageref{firstpage}--\pageref{lastpage}} \pubyear{2011}

\maketitle

\label{firstpage}

\begin{abstract}
We report on the \swift\ detection of a thermonuclear X-ray burst from the very-faint quasi-persistent neutron star X-ray binary \xmmuname, which triggered the satellite's Burst Alert Telescope (BAT) on 2010 August 13. Analysis of the BAT spectrum yields an observed bolometric peak flux of $\simeq 4.5 \times 10^{-8}~\flux$, from which we infer a source distance of $\lesssim 8.4$~kpc. Follow-up observations with \swift's X-ray Telescope (XRT) suggest that the event had a duration of $\simeq 3$~h and a total radiated energy of $\simeq9\times10^{40}$~erg, which classify it as an intermediately long X-ray burst. This is only the second X-ray burst ever reported from this source. Inspection of \swift/XRT observations performed between 2007--2010 suggests that the 2--10 keV accretion luminosity of the system is $\simeq5 \times 10^{34}~\lum$ for an assumed distance of 8.4~kpc. Despite being transient, \xmmuname\ appears to have been continuously active since its discovery in 2003. 
\end{abstract}

\begin{keywords}
accretion, accretion discs -- stars: neutron -- X-rays: binaries -- X-rays: bursts -- X-rays: individual (\xmmuname, \igr)
\end{keywords}

\section{Introduction}\label{sec:intro}
In neutron star low-mass X-ray binaries (LMXBs), a star of (sub-) solar mass is feeding matter to the compact primary via an accretion disk. This typically gives rise to a 2--10 keV accretion luminosity of $L_{\mathrm{X}}\sim10^{36-38}~\lum$. However, a small group of LMXBs, the very-faint systems, display sub-luminous accretion intensities of $L_{\mathrm{X}}\sim10^{34-36}~\lum$ \citep[][]{wijnands06}. Many LMXBs are transient and alternate accretion outbursts with episodes of quiescence, during which the X-ray luminosity is orders of magnitude lower.

Unstable thermonuclear burning of helium (He) and/or hydrogen (H) accreted onto the surface of a neutron star, results in a type-I X-ray burst (shortly `X-ray burst' hereafter). These can temporarily outshine the accretion luminosity and are characterized by blackbody emission with a peak temperature of $kT_{\mathrm{bb}}\simeq2-3$~keV. Typically, a fast rise is followed by a slower decay, during which the blackbody temperature decreases. Some X-ray bursts show photospheric radius expansion (PRE), as evidenced by a local peak in emitting radius associated with a drop in blackbody temperature. These are thought to reach the Eddington limit, allowing for a distance determination \citep[][]{kuulkers2003}.

The properties (e.g., duration, radiated energy and recurrence time) of X-ray bursts depend on the conditions of the ignition layer, such as the temperature, thickness and H abundance. These are sensitive to the mass-accretion rate onto the neutron star, and consequently the characteristics of X-ray bursts depend on the accretion regime \citep[e.g.,][]{fujimoto81}. 
Most observed X-ray bursts last for $\sim10-100$~s, release a total energy of $E_{\mathrm{burst}}\sim10^{39}~\mathrm{erg}$, and recur every few hours to days \citep[e.g.,][]{galloway06}. 

Much more rare are the intermediately long X-ray bursts, which can last for tens of minutes, have a total radiated energy of $E_{\mathrm{burst}}\sim10^{40-41}~\mathrm{erg}$, and are thought to recur only after many days. These are likely caused by the ignition of a thick layer of He, and are detected from neutron stars that are thought to have a relatively cold envelope, either because they accrete matter only slowly \citep[$\lesssim1\%$ of the Eddington rate;][]{cooper07}, or due to the fact that the accreted matter is H-poor and hence heating from H-burning is absent \citep[][]{cumming06}. Roughly $10\%$ of all bursting neutron star LMXBs exhibit intermediately long X-ray bursts \citep[e.g.,][]{falanga08}.

\subsection{\xmmuname}
The X-ray source \xmmuname\ (\xmmu\ hereafter) was serendipitously discovered during \xmm\ observations of the supernova remnant G0.9+0.1, performed on 2003 March 12 \citep[][]{sidoli2003}. The source displayed a 2--10 keV luminosity of $L_{\mathrm{X}}\simeq5\times10^{34}~\lum$ (assuming a distance $D=8.4$~kpc; see Sec.~\ref{subsec:bat}), but was not detected during previous \xmm\ observations of the field obtained on 2000 September 23 \citep[][]{sidoli2003}. This implied that the source flux was variable by at least a factor of $\sim100$, demonstrating its transient nature. A search through archival data showed that the source was not detected during \chan\ observations carried out on 2000 October 27 and 2001 July 16, resulting in an upper limit on the quiescent luminosity of $L_{\mathrm{q}}\lesssim 10^{32}~\lum$ \citep[2--10 keV;][]{delsanto07}. The low outburst intensity classifies \xmmu\ as a very-faint X-ray transient.

On 2005 March 22, \inte\ detected an X-ray burst from a source designated \igr\ \citep[][]{brandt2006_xmmu}. As argued by \citet{wijnands2006_xmmu}, the bursting source was likely associated with \xmmu, establishing its nature as a neutron star LMXB. Based on the properties of the X-ray burst, \citet{delsanto07} argued that the source was likely continuously active between 2003 and 2005, and was thus undergoing a very long accretion outburst. This idea was strengthened by the fact that \xmmu\ was detected in outburst every time that \xmm, \chan\ or \swift\ covered the source field since 2003 \citep[][]{campana09,degenaar07_xmmsource,degenaar2007_xmmsource2,delsanto07,delsanto2009,delsanto2010,sidoli2007}. Systems that undergo such unusually long outburst episodes are denoted as quasi-persistent X-ray transients. 

On 2010 August 13 at 21:03 \textsc{ut}, \swift's Burst Alert Telescope (BAT) triggered on a source located near the Galactic centre, consistent with the position of \xmmu\ \citep[trigger 431582;][]{gelbord2010}. In this Letter, we analyze the BAT data of this event, as well as the XRT follow-up observation, demonstrating that this trigger was caused by a long X-ray burst from \xmmu. We use archival \swift/XRT data, obtained between 2007 and 2010, to characterize the persistent emission of the source.

\vspace{-0.2cm}

\section{Observations, analysis and results}\label{sec:data}

\subsection{\swift/BAT}\label{subsec:bat}
\begin{figure}
 \begin{center}
         \includegraphics[width=8.0cm]{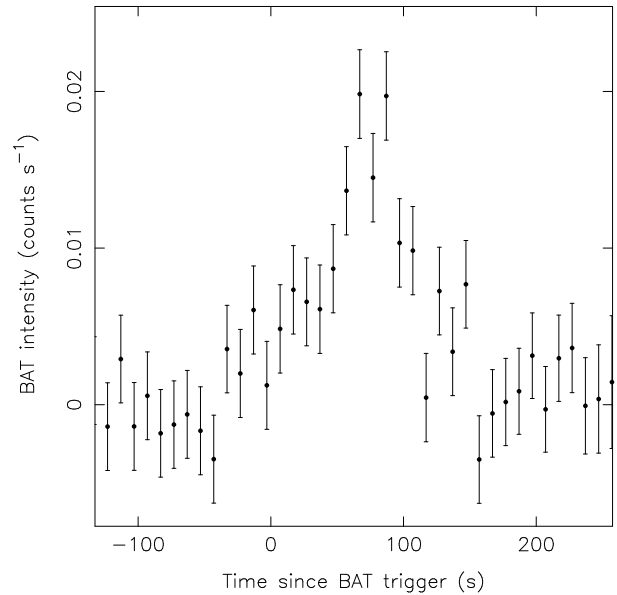}
    \end{center}
\caption[]{{
\swift/BAT lightcurve at 10-s resolution (15--35 keV).
}}
 \label{fig:bat}
\end{figure}

We generated standard BAT data products using the \textsc{batgrbproduct} tool. The 15--35 keV BAT lightcurve, shown in Fig.~\ref{fig:bat}, is consistent with a single peak centred at $t\simeq75$~s and emerging from the background for $\simeq 150$~s, with a very slow rise time of $\simeq 100$~s. The spacecraft started slewing $\simeq200$~s after the burst trigger, by which time the BAT count rate had returned to the background level (see Fig.~\ref{fig:bat}). Limited by a low number of photons, we extracted a single spectrum of 150 s around the burst peak, using the tool \textsc{batbinevt}. We applied the necessary geometrical corrections with \textsc{batupdatephakw}, administered the BAT-recommended systematical error using \textsc{batphasyserr}, and generated a single response matrix by running the task \textsc{batdrmgen}.

\begin{figure}
 \begin{center}
         \includegraphics[width=8.0cm]{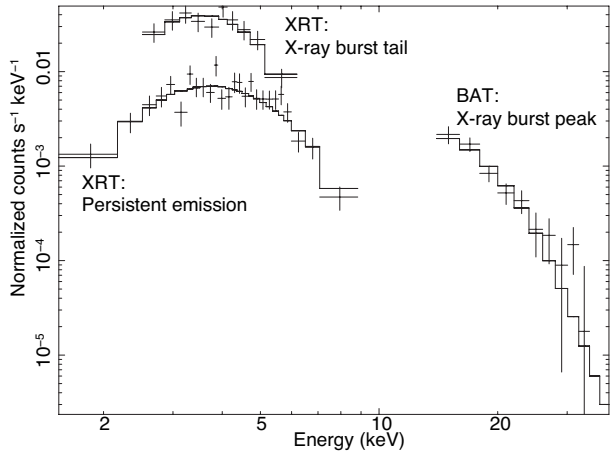}
    \end{center}
\caption[]{{
Spectra of the X-ray burst seen by BAT (right) and XRT (top left), as well as the persistent emission of \xmmu\ (bottom left). The solid lines indicate the model fits (see text). 
}}
 \label{fig:spec}
\end{figure}

The BAT spectrum is relatively soft, with no photons detected above $\simeq35$~keV (see Fig.~\ref{fig:spec}). Fitting the data between 15--35 keV with \textsc{Xspec} (v. 12.6) to a simple powerlaw continuum yields a photon index of $\Gamma=5.8\pm0.8$ for a reduced chi-squared value of $\chi^2_{\nu}=0.95$ and 10 degrees of freedom (dof). Such a high spectral index is suggestive of a thermal shape. A blackbody (\textsc{bbodyrad}) model fit, shown in Figure~\ref{fig:spec}, results in a temperature of $kT_{\mathrm{bb}}=2.45\pm0.40$~keV and an emitting radius of $R_{\mathrm{bb}}=6.3^{+3.5}_{-1.0}$~km (for $D=8.4$~kpc; see below), yielding $\chi^2_{\nu}=1.08$ for 10 dof. This suggests that the BAT triggered on an X-ray burst.

We extrapolate the blackbody fit to the 0.01--100 keV energy range and estimate an unabsorbed bolometric flux of $F_{\mathrm{bol}}^{\mathrm{BAT}}=2.3^{+2.0}_{-0.9} \times10^{-8}~\flux$, which gives a fluence for the 150-s interval of $f_{\mathrm{BAT}}=3.5\times10^{-6}~\fluence$. 
The average BAT count rate during this interval is $1.20\times10^{-2}~\cnts$, whereas the observed peak count rate is $2.34\times10^{-2}~\cnts$. We thus expect that the peak flux of the X-ray burst lies a factor of $1.95$ higher, at $F_{\mathrm{bol}}^{\mathrm{peak}}\simeq4.5\times10^{-8}~\flux$. This is similar to the peak flux measured for the X-ray burst detected from \xmmu\ with \inte\ on 2005 March 22 \citep[][]{delsanto2007}.

Due to the low number of counts and the large gap with the XRT follow-up observation (see Sec.~\ref{subsubsec:trigger}), we cannot investigate whether this X-ray burst exhibited a PRE phase. However, by equating the BAT peak intensity to the luminosity typically observed from PRE X-ray bursts \citep[$L_{\mathrm{edd}}=3.8\times10^{38}~\lum$;][]{kuulkers2003}, we can constrain the source distance to be $D\lesssim8.4$~kpc. This is consistent with the value inferred from the X-ray burst of \xmmu\ detected by \inte\ \citep[][]{delsanto2007}. 
For a photosphere of solar composition rather than pure He, the Eddington luminosity is $L_{\mathrm{edd}}\simeq1.5~\times10^{38}~\lum$, which would lower the distance estimate to $D\lesssim5.3$~kpc. Throughout this Letter, we have adopted a source distance of $D=8.4$~kpc.

\subsection{\swift/XRT}\label{subsec:xrt}

To obtain cleaned data products, we processed raw XRT data with the task \textsc{xrtpipeline}, using standard quality cuts and selecting event grades 0--12. Source lightcurves and spectra were extracted with \textsc{xselect} (v. 2.4). We collected source photons from a circular region with a 15-pixel radius centred around \xmmu. The background emission was averaged over a set of three circular regions of similar dimensions, which were placed on nearby, source-free parts of the CCD. We generated exposure maps with the task \textsc{xrtexpomap} and ancillary response files (arf) were created with \textsc{xrtmkarf}. The latest response matrix files (v. 11; rmf) were obtained from the CALDB database. The XRT spectra were grouped to a minimum of 20 photons per bin and fitted in the 0.5--10 keV energy range. For the interstellar hydrogen absorption we employ the \textsc{phabs} model.

\subsubsection{Persistent emission}\label{subsubsec:persistent}
We used all publicly available \swift/XRT observations of \xmmu\ to characterize its persistent emission. This concerns a total of 10 individual exposures, varying between 1 and 4 ks in length, and spanning the time between 2007 May 13 and 2010 November 2. The total exposure time is $\simeq20.5$~ks. \xmmu\ is detected at an average count rate of $2.53\times10^{-2}~\cnts$, varying by a factor $\sim2$. Most pointings did not collect sufficient photons to fit each spectrum individually, so we summed all observations.

The composite X-ray spectrum, displayed in Figure~\ref{fig:spec}, can be described by an absorbed powerlaw model with $N_H=(8.6\pm2.3)\times10^{22}~\nh$ and $\Gamma=2.2\pm0.5$ ($\chi^2_{\nu}=1.14$ for 23 dof). These spectral parameters are similar to the values obtained from \xmm\ data \citep[$N_H=8.9\pm0.5\times10^{22}~\nh$ and $\Gamma=2.1\pm0.1$;][]{delsanto07}. The corresponding absorbed and unabsorbed 2--10 keV fluxes are $F\mathrm{_{X}^{abs}}=3.4\times10^{-12}~\flux$ and $F\mathrm{_{X}^{unabs}}=6.4\times10^{-12}~\flux$, respectively. For a distance of 8.4~kpc, the latter corresponds to a luminosity of $L_{\mathrm{X}}=5.4\times10^{34}~\lum$. If the bolometric luminosity is a factor $\sim3$ higher than measured in the 2--10 keV band \citep[][]{zand07}, this gives $L^{\mathrm{pers}}_{\mathrm{bol}}\simeq1.6\times10^{35}~(D/8.4~\mathrm{kpc})^2~\lum$.

\subsubsection{X-ray burst}\label{subsubsec:trigger}
In general, the \swift\ spacecraft slews towards the position of a BAT trigger, allowing for follow-up observations with the narrow-field instruments, within a few minutes. However, in case of trigger 431582, the Earth limb observing constraint delayed the slewing \citep{gelbord2010}. The XRT follow-up observation commenced on 2010 August 13 at 21:58 \textsc{ut}, approximately 55~min after the BAT trigger. 

\begin{figure}
 \begin{center}
         \includegraphics[width=8.0cm]{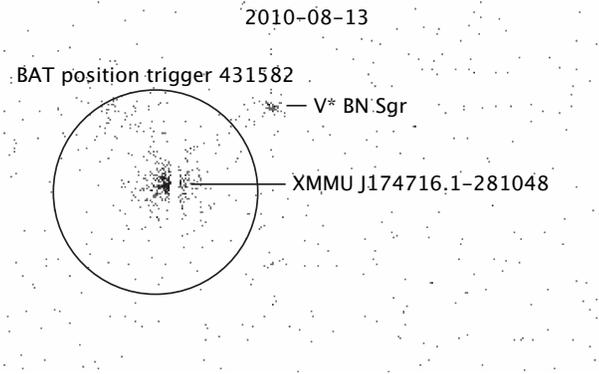}
    \end{center}
\caption[]{{\swift/XRT image (0.3--10 keV) obtained $\simeq1$~h after the BAT trigger. The $2.4'$ BAT error circle is indicated.
}}
 \label{fig:ds9}
\end{figure}

The XRT observation had an exposure time of $\simeq2.2$~ks, during which the instrument was operated in the photon counting (pc) mode. In the 0.3--10 keV XRT image, shown in Figure~\ref{fig:ds9}, \xmmu\ is the only X-ray point source located within the $\simeq 2.4'$ BAT position uncertainty. A second X-ray point source is visible, which lies $\simeq 3.4'$ from the BAT coordinates at a position that is consistent with V* BN Sgr, an eclipsing binary of Algol type. However, we show below that it was likely \xmmu\ that caused the BAT trigger, as was also suggested by \citet{gelbord2010}.

The spectrum of \xmmu\ obtained during this observation, displayed in Figure~\ref{fig:spec}, can be fit with an absorbed powerlaw model with a spectral index $\Gamma=3.2\pm0.5$, yielding $\chi^2_{\nu}=0.92$ (8 dof), for a fixed hydrogen column density of $N_H=8.6\times10^{22}~\nh$ (see Sec.~\ref{subsubsec:persistent}). The photon index is softer than obtained for other XRT observations of \xmmu\ ($\Gamma=2.2\pm0.5$; see Sec.~\ref{subsubsec:persistent}), and suggests that the spectrum rather has a thermal shape. For an absorbed blackbody model, we obtain $kT_{\mathrm{bb}}=0.78\pm0.09$~keV and $R_{\mathrm{bb}}=3.9^{+1.3}_{-0.8}$~km ($\chi^2_{\nu}=0.6$ for 8 dof). The inferred blackbody temperature is consistent with the tail of an X-ray burst \citep[e.g.,][]{linares09,degenaar2010_burst}. By extrapolating the fit to the 0.01--100 keV energy range, we obtain an estimated bolometric flux of $F_{\mathrm{bol}}^{\mathrm{XRT}}=8.7^{+3.4}_{-1.9} \times10^{-11}~\flux$.

In order to investigate whether the data exhibits the characteristic spectral softening of X-ray bursts, we split the XRT observation into two separate intervals of similar length. We extracted spectra for the two data segments and fitted these to an absorbed blackbody model. The hydrogen column density was fixed at $N_H=8.6\times10^{22}~\nh$, and the emitting area was assumed to remain constant. This yielded blackbody temperatures of $kT_{\mathrm{bb}}=0.88\pm0.09$ and $0.80\pm0.08$~keV for the first and second interval, respectively. Although the errors are substantial due to the low number of counts in the spectra ($\simeq 150$ per interval), it is suggestive of cooling along the decay. This further strengthens the idea that \xmmu\ was exhibiting an X-ray burst.

\begin{figure}
 \begin{center}
         \includegraphics[width=8.0cm]{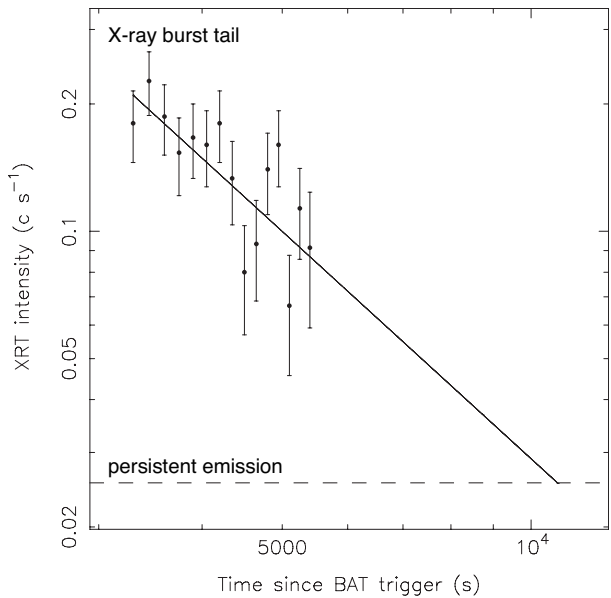}
    \end{center}
\caption[]{{
\swift/XRT lightcurve of \xmmu\ observed on 2010 August 13, at 150-s time resolution (0.3--10 keV). The dashed line indicates the persistent emission level and the solid line represents a fit to a powerlaw decay with index $-1.8$. 
}}
 \label{fig:xrt}
\end{figure}

During the XRT observation of 2010 August 13, \xmmu\ is detected at an average count rate of $\simeq0.1~\cnts$, which is a factor $\sim5-10$ elevated above its typical emission level (see Sec.~\ref{subsubsec:persistent}). As shown in Figure~\ref{fig:xrt}, \xmmu\ shows a clear decay in count rate during the 2.2-ks long XRT exposure. We fitted the lightcurve, binned by 150-s, to a powerlaw decay, which resulted in an index of $-1.8\pm 0.4$ and a normalization of $\simeq4\times10^{5}~\cnts$ ($\chi^2_{\nu}=1.10$ for 13 dof). The value of the decay index is consistent with the theoretical prediction for the cooling tail of a long X-ray burst \citep[][]{cumming04}, and similar to observational results of intermediately long X-ray bursts from other sources \citep[e.g.,][]{falanga08,linares09,degenaar2010_burst}. 

The total length of the X-ray burst can be determined by extrapolating the above described powerlaw decay down to the persistent emission level. This suggests a duration of $t_{\mathrm{burst}}\simeq10500$~s ($\simeq2.9$~h; see Fig.~\ref{fig:xrt}). To estimate the fluence in the X-ray burst tail, we integrate the powerlaw decay function from $t=150$~s (the time at which the burst peak disappeared from the BAT lightcurve; see Fig.~\ref{fig:bat}) till $t=10500$~s after the BAT trigger. By applying a count rate to flux conversion factor inferred from fitting the XRT spectrum of the X-ray burst, we find a bolometric fluence of $f_{\mathrm{XRT}}\simeq 7.3 \times 10^{-6}~\fluence$. Adding this to the BAT result (see Sec.~\ref{subsec:bat}), the total estimated bolometric fluence of the X-ray burst becomes $f_{\mathrm{burst}}\simeq 1.1 \times 10^{-5}~\fluence$. For a distance of $D=8.4$~kpc, this implies a radiated energy of $E_{\mathrm{burst}} \simeq 9.1 \times 10^{40}$~erg. The ignition column depth of an X-ray burst is given by $y=E_{\mathrm{burst}}(1+z)/4 \pi R^2 Q_{\mathrm{nuc}}$, where $z$ is gravitational redshift, $R$ is the neutron star radius and $Q_{\mathrm{nuc}}=1.6+4X~\mathrm{MeV~nucleon}^{-1}$, the nuclear energy release given a H-fraction $X$ at ignition \citep[e.g.,][]{galloway06}. For a neutron star with $M=1.4~\Msun$ and $R=10$~km ($z=0.31$), we find $y\simeq 6.2 \times 10^{9}~\mathrm{g~cm}^{-2}$ for pure He ($X=0$), or $y\simeq 2.3 \times 10^{9}~\mathrm{g~cm}^{-2}$ for solar abundances ($X=0.7$). 

The recurrence time that corresponds to a certain ignition depth is $t_{\mathrm{rec}} \simeq y(1+z)/\dot{m}$, where $\dot{m}$ is the accretion rate onto the neutron star surface per unit area. The bolometric persistent luminosity of \xmmu\ is $L_{\mathrm{bol}}^{\mathrm{pers}}\simeq 1.6 \times 10^{35}~(D/8.4~\mathrm{kpc})^2~\lum$ (see Sec.~\ref{subsubsec:persistent}), which yields a mass-accretion rate of $\dot{M} \simeq R L_{\mathrm{bol}}^{\mathrm{pers}}/GM \simeq 1.4 \times 10^{-11}~\Msun~\mathrm{yr}^{-1}$ ($\simeq0.05\%$ of Eddington). If the emission is isotropic, this corresponds to a local accretion rate of $\dot{m}\simeq 68~\mathrm{g~cm}^{-2}~\mathrm{s}^{-1}$. We can thus roughly estimate that the time required to build up the layer that caused the X-ray burst is $t_{\mathrm{rec}} \simeq 3.8$~yr ($X=0$) or $t_{\mathrm{rec}}\simeq1.4$~yr ($X=0.7$). 

\section{Discussion}\label{sec:discuss}
In this Letter, we analyzed the \swift\ data of BAT trigger 431582, which occurred on 2010 August 13. The BAT spectrum fits to a blackbody model with a temperature of $kT_{\mathrm{bb}}^{\infty}\simeq2.5$~keV and an emitting radius of $R_{\mathrm{bb}}\simeq6.3$~km, suggesting that this was a thermonuclear event. Follow-up observations with \swift/XRT, performed $\simeq1$~h after the BAT trigger, reveal that the known X-ray burster \xmmuname\ was the only X-ray point source located with the $\simeq2.5'$ BAT error circle. Analysis of the XRT data shows that the intensity of \xmmu\ was elevated above its persistent level by a factor of $\sim5-10$, while the source lightcurve displayed a clear decaying trend that fits to a powerlaw with index $-1.8$.  
This provides strong evidence pointing towards \xmmu\ as the source of the BAT trigger. The fact that the XRT spectrum can be described by a blackbody model with a temperature of $kT_{\mathrm{bb}}^{\infty}\simeq0.8$~keV, and shows indications of cooling during the decay, 
further strengthens the idea that the BAT triggered on an X-ray burst from \xmmu. 

From the observed BAT peak flux, we derive an upper limit on the distance towards \xmmu\ of $D\lesssim8.4$~kpc. A location near the Galactic centre is consistent with the large hydrogen column density inferred from fitting the persistent X-ray spectrum ($N_H\simeq9\times10^{22}~\nh$). The total radiated energy of $E_{\mathrm{burst}}\simeq 9\times10^{40}$~erg, classifies this event as an intermediately long X-ray burst \citep[e.g.,][]{falanga08,zand08,degenaar2010_burst}. The fact that the X-ray burst could be observed for a very long time ($\simeq 1.5$~h) and extrapolated to a duration of $\simeq 2.9$~h, can be attributed to the low persistent emission level of the source, combined with the sensitivity and low X-ray background of the \swift/XRT. This provides the unique possibility to follow the tails of X-ray bursts down to very low temperatures below $\sim1$~keV. The inferred ignition column depth ($y\simeq 2-6\times10^{9}~\mathrm{g~cm}^{-2}$) requires that \xmmu\ accreted for $\sim1-4$ years to be able to power this X-ray burst. 

Investigation of \swift/XRT data obtained between 2007 May and 2010 November, reveals that \xmmu\ was likely continuously active, displaying a 2--10 keV luminosity of $L_{\mathrm{X}}\simeq5\times10^{34}~(D/8.4~\mathrm{kpc})^2~\lum$. Similar intensities were measured with \xmm\ in 2003 March and 2005 February \citep[][]{delsanto07}, and with \chan\ in 2007 May \citep[][]{degenaar2007_xmmsource2}. This supports the suggestion of \citet{delsanto07} and \citet{degenaar07_xmmsource}, that the system exhibits a very long accretion outburst. The detection history suggests that the outburst started between 2001 July 16 and 2003 March 12. This sets the counter on the outburst duration to 7--9 years. The mass-accretion rate ($\simeq0.05\%$ of Eddington) is amongst the lowest encountered for currently known neutron star LMXBs.

The X-ray burst observed from \xmmu\ with \swift\ is only the second one ever reported from this source. The first X-ray burst was detected with \inte\ on 2005 March 22, and reached a peak flux of $F_{\mathrm{bol}}^{\mathrm{peak}}\simeq5 \times10^{-8}~\flux$ \citep[][]{delsanto2007}, similar to our result for the \swift\ event. The \inte\ X-ray burst was detected with the IBIS/ISGRI in the 18--26 energy band for $\simeq100$~s \citep[][]{delsanto07}, comparable with the visibility in the 15--35 BAT lightcurve. In the 3--6 keV band, JEM-X detected the X-ray burst for $\simeq200$~s, i.e., much shorter than observed with \swift/XRT. However, JEM-X has a lower sensitivity and no energy coverage below $\simeq3$~keV, where the tail of the X-ray burst emission lies. The total radiated energy of the \inte\ event, $E_{\mathrm{burst}}\simeq2\times10^{40}$~erg \citep[][]{delsanto07}, suggests that this was likely an intermediately long X-ray burst as well. From the energy budget of the \swift\ X-ray burst, we estimate a recurrence time of $\sim 1-4$~years, depending on the photospheric composition. For pure He, this is roughly consistent with the time elapsed since the \inte\ X-ray burst (5 years). However, additional X-ray bursts might have been missed.

There are two scenarios for which intermediately long X-ray bursts may occur. First, in ultra-compact X-ray binaries (UCXBs), the mass donor is H-poor, so that the neutron star is accreting nearly pure He. In absence of heating from H-burning, the temperature in the ignition layer is low. As a consequence, a thick layer of He can build up before it ignites, giving rise to a long and energetic X-ray burst \citep[][]{cumming06}. Second, some sources accrete matter from a H-rich companion at such a low rate that H burns unstably. For these very-faint LMXBs, successive weak H-flashes can result in the production of a large reservoir of He that also causes an intermediately long X-ray burst upon ignition \citep[][]{cooper07,peng2007}. There are observational examples for both scenarios \citep[e.g.,][]{zand05,chenevez07,falanga08,falanga09,degenaar2010_burst,kuulkers09}. 

Without a direct measurement of the orbital period, clues to the ultra-compact nature of an X-ray binary may come from a comparison of the X-ray and optical flux \citep[e.g.,][]{bassa06}, or abundance patterns revealed through optical spectroscopy \citep[e.g.,][]{nelemans2010}. However, the inferred large hydrogen column density of \xmmu\ ($N_H\simeq9\times10^{22}~\nh$) implies a visual extinction of $\simeq50$~mag \citep[][]{predehl1995}, rendering searches for an optical counterpart basically impossible. Although \xmmu\ is a confirmed transient system, it appears able to keep the accretion ongoing for many years. This suggests the possibility of a relatively small orbit, and UCXB nature, since small accretion disks are easier to be kept fully ionized so that the accretion can be sustained \citep[][]{zand07}. However, the neutron star LMXB \rxh\ is likely persistently accreting at $L_{\mathrm{bol}}^{\mathrm{pers}}\simeq4 \times10^{35}~\lum$, but shows a strong $\mathrm{H\alpha}$ emission line that indicates a H-rich donor \citep[][]{degenaar2010_burst}.

Intermediately long X-ray bursts that are likely triggered by unstable H-burning have been observed at mass-accretion rates of $\simeq0.1-1\%$ of Eddington \citep[e.g.,][]{degenaar2010_burst,linares09,zand08}. Initial calculations predict that these should occur only for a narrow regime spanning a factor $\simeq3$ in mass-accretion rate \citep[][]{cooper07,peng2007}. In contrast, if \xmmu\ (accreting at $\simeq0.05\%$ Eddington) fits into this framework, the observations would span a factor $\simeq20$ in mass-accretion rate. However, the theoretically allowed range relaxes if the heat flux coming from the neutron star crust varies between the different sources \citep[][]{cooper07}, and this is indeed expected to depend on the mass-accretion rate onto the neutron star \citep[][]{cumming2000,brown2004}. So far, it thus remains unclear which of the two possibilities sketched above apply to the intermediately long X-ray burst observed from \xmmu\ with \swift. \\

\noindent
\textbf{Acknowledgments}\\
This work was supported by the Netherlands Research School for Astronomy (NOVA) and makes use of the \swift\ public data archive. RW and RK acknowledge support from a European Research Council (ERC) starting grant. The authors are grateful to the referee, J\'{e}r$\mathrm{\hat{o}}$me Chenevez, for useful comments.

\vspace{-0.5cm}
\bibliographystyle{mn2e}

\label{lastpage}
\end{document}